\documentclass{aa}
\usepackage{epsfig}
\usepackage{aalongtable}
\usepackage{txfonts}

\newcommand{\teff}{T$_{\rm eff}$}
\newcommand{\nli}{$\log$~n(Li)}
\newcommand{\nbe}{$\log$~n(Be)}

\newcommand{\beii}{\ion{Be}{ii}}
\newcommand{\kms}{\mbox{\rm km\,s$^{-1}$}}

\begin{document}

\title{Tracing mixing in stars: new beryllium observations of
the open clusters NGC~2516, Hyades, and M~67
\thanks{Based on observations collected at ESO-VLT, Paranal Observatory,
Chile, Programme numbers 65.L-0427, 68.D-0491, 69.D-0454}}

   \subtitle{}

   \author{S. Randich\inst{1} \and F. Primas\inst{2} 
         \and L. Pasquini\inst{2} \and P. Sestito\inst{1} 
\and R. Pallavicini\inst{3}}

   \offprints{S. Randich, email:randich@arcetri.astro.it}

\institute{INAF/Osservatorio Astrofisico di Arcetri, Largo E. Fermi 5,
             I-50125 Firenze, Italy
\and
European Southern Observatory, Garching bei M\"unchen, Germany
\and
INAF/Osservatorio Astronomico di Palermo, Piazza del
                  Parlamento 1, I-90134 Palermo, Italy}

\titlerunning{Beryllium in open clusters}
\date{Received Date: Accepted Date}


\abstract
{
Determinations of beryllium abundance in stars, together with lithium,
provide a key tool to investigate the so far poorly understood
extra-mixing processes at work in stellar interiors.
}
{We measured Be in
three open clusters, complementing existing Be surveys, and aiming 
at gathering a more complete empirical scenario of
the evolution of Be as a function of stellar age and temperature.
}
{We analyzed VLT/UVES spectra of members of \object{NGC~2516}, the 
\object{Hyades}, and
\object{M~67} to determine their Be and Li abundances.
In the first two clusters we focused on stars cooler than 5400~K,
while the M~67 sample includes stars warmer than
6150~K, as well as two subgiants and two blue stragglers.
We also computed the evolution of Be for a 0.9~M$_{\odot}$ star based on
standard evolutionary models.
}
{We find different behaviours for stars in different temperature
bins and ages. Stars warmer than 6150~K show Be depletion and 
follow a Be vs. Li correlation, while Be is undepleted in
stars in the $\sim 6150-5600$~K range.
NGC~2516 members cooler than 5400~K have not depleted any Be, while
older Hyades of similar temperature show some depletion.
Be is severely depleted in the subgiants and blue stragglers.
}
{The results for warm stars
are in agreement with those of previous studies, supporting
the hypothesis that
mixing in this temperature regime is driven by rotation. The same
holds for the two subgiants that have evolved from the ``Li gap".
This mechanism is instead not the dominant one for solar-type stars.
Be depletion of cool Hyades cannot
simply be explained by the effect of increasing depth of the convective zone.
Finally,
the different Be content of the two blue stragglers
suggests that they have formed by two different processes (i.e.,
collisions vs. binary merging).
}

\authorrunning{S. Randich et al.}
\maketitle
\section{Introduction}\label{intro}
Beryllium (Be) is easily burned in  stellar interiors at
temperatures  above $\sim$ 3.5 MK; as a consequence,
any process that is able to mix
surface material with internal Be-depleted material results 
in surface Be depletion. Since the more fragile lithium (Li) 
is destroyed at temperatures above $\sim$ 
2.5 MK, measurements of both elements in the same star allow us
to investigate how deep mixing extends in the stellar interior
and to put constraints on the various extra-mixing processes
introduced in evolutionary models to explain the complex
Li behavior.
Several unexpected results related to the depletion of Li in stars
in fact have been found, 
which reveal our limited understanding of the physics 
acting in the interiors of stars. 

Focusing on late-F and early G-type stars, different empirical evidences 
contradict the predictions of standard 
models (those including convection only). In particular, observations
of Li in open clusters of different ages have
shown that these stars deplete Li during the main sequence (MS), in spite of
the fact that their convective zones do not reach deep enough in the stellar
interior to meet the zone of Li destruction
(e.g., Randich \cite{r_cast} and references therein);
the factor of $\sim 100$ Li depletion measured in  the Sun itself
indeed represents a long standing puzzle. Moreover,
a factor of $\sim 10$ scatter in Li is observed in the solar-age,
solar metallicity cluster M~67
(Spite et al. \cite{spi87}; Garc\'\i a L\'opez et al. \cite{gar88};
Pasquini et al.~\cite{pas97}; Jones et al.~\cite{jon99}), indicating that
otherwise similar stars 
(i.e., stars with the same age, temperature, and metallicity)
may be affected by different amounts of Li depletion, again
in disagreement with standard models. 
This dispersion implies that Li depletion must be affected by
an additional parameter besides mass, age, and chemical composition.

As to warmer early--F stars,
since the 80s' Boesgaard and collaborators 
pointed out the presence of a so-called ``Li gap"  or ``Li dip"
among hot MS stars in the Hyades cluster, 
namely a sharp decrease of Li in a narrow ($\sim 300$~K) effective
temperature (\teff) range around 6600~K (Boesgaard \& Tripicco \cite{bt86}).
The gap is present in several other clusters of the Hyades age or 
older (Balachandran~\cite{bala95} and references therein), is weak
or absent in the 120 Myr \object{Pleiades} (Pilachowski et al.~\cite{piletal87};
Boesgaard et al.~\cite{boesetal88}), but has been detected in the slightly older
\object{M~35} (160~Myr --Steinhauer \& Deliyannis \cite{std04}), suggesting that
it develops at ages below $\sim 200$~Myr.
 
In order to explain the unexpected MS Li depletion of F and early-G type stars,
a number of models and extra-mixing processes have been proposed, 
which include mass loss (Swenson \& Faulkner \cite{faulk}),
diffusion (Michaud \cite{michaud}; Chaboyer et al. \cite{chab}),
meridional circulation (Charbonnel \& Talon \cite{ct99} and references
therein), angular momentum 
loss, and rotationally driven mixing (Schatzman \& Baglin~\cite{sb91};
Pinsonneault et al.~\cite
{pins92}; Deliyannis \& Pinsonneault \cite{dp97}), 
gravity waves (Garc\'\i a L\'opez \& Spruit \cite{gs91};
Montalban \& Schatzmann \cite{ms00}), 
tachocline (Brun et al. \cite{brun}; Piau et al. \cite {piau}),  
and combinations of these (Charbonnel \& Talon \cite{ct05}). 
Note that models including rotational mixing would in principle
be able to explain the scatter il Li in M~67, with initial angular momentum
being the additional parameter affecting Li depletion.
Since models including different extra-mixing processes predict
distinct behaviours of Li vs. Be with age and with mass, 
the last few years have witnessed a flourishing of Be observations
among field and cluster stars. The availability of state-of-the art
high resolution spectrographs with high near-UV efficiency
has made it possible to observe not only bright stars in close-by clusters,
but also fainter members of more distant, old clusters.

These Be surveys have allowed constraining the mechanism
responsible for the gap; more specifically,
several studies have now evidenced the presence of a
correlation between Li and Be among cluster and field stars
on the cool/red side of the gap (\teff~between $\sim 6600$ and 6300~K),
providing strong support to rotationally induced mixing 
(e.g., Deliyannis et al. \cite{deli98}; Boesgaard et al.
\cite{bk04}, \cite{boes04}).
As for cooler stars in older clusters, 
Randich et al. (\cite{R02}; hereafter R02) found that
solar-type members (\teff~$\sim 6150-5800$~K) of M~67 and the 
2~Gyr old IC~4651 have not depleted any Be
and thus do not follow any Li-Be correlation.
This finding allowed them to rule out a number of models;
in particular, at variance with
warmer stars on the red side of the gap, present
models including rotationally induced mixing do not seem to well
reproduce the
observed Be vs. Li pattern for stars cooler than $6150$~K in M~67:
whereas they predict that at old ages these stars should have depleted
some Be (a factor 2--3), no depletion is instead observed.

In this work we present new Be measurements
in three open clusters: the young NGC~2516
(150 Myr), the Hyades (600 Myr), and the old M~67 (4.5 Gyr). 
In NGC~2516 and the Hyades we mostly concentrated 
on cool MS stars (\teff~below 5400 K), the reason being that 
warmer stars in both the Hyades and other young clusters (such as the Pleiades
and Alpha Per) have been extensively studied in the past 
by Boesgaard and collaborators (see Boesgaard et al.~\cite{boes04}
and references therein).
In M~67 we instead observed stars warmer
than in the R02 sample, as well as two subgiants and two blue stragglers.
Since the upper turnoff of M~67 coincides with the cool 
edge of the Li gap, subgiants 
are supposed to be evolved from objects that were in the gap (Balachandran
\cite{bala95}); their Li and Be abundances 
are therefore important to constrain the gap evolution
at old ages. As to blue stragglers, their light element content might provide
insights into their formation mechanism.
In summary, when joining our sample with available Be cluster data 
from the literature, we have a coverage of the \teff~interval
from the Li gap down to 5100~K
for three groups of stars, with ages of 0.15, 0.6 and 4.5~Gyr. 
Based on this sample, our goal is to obtain a more complete
empirical picture of the evolution of Be (and Be vs. Li) as a function
of stellar age and temperature.
\section{Sample stars and observations}
The sample includes
four members of NGC~2516, four members of the Hyades and
eight members of M~67.
%
The observations were carried out in Service mode using UVES on VLT/Kueyen
(Dekker et al.~\cite{dek}) during the period November 2001 -- March 2002.
All stars were observed using the same set-up; namely,
UVES was operated in Dichroic Mode with Cross Dispersers \#1 and \#3 in the 
Blue and Red arms, respectively. 
Such a combination allowed us to
cover the spectral ranges from $\sim 3115 $ to 3940~\AA~in the blue and
from $\sim$ 4780 to 6810~\AA~in the red. The 15$\mu m$ pixels
and the use of a 1~arcsec wide
slit (projecting onto 4 pixels) and CCD binning $2\times 2$ and 
$1\times 1$ in the blue and red yielded
resolving powers R$\sim 40,000$ and R$\sim 45,000$, respectively.
The spectra were reduced by the ESO Quality Control Group using the
UVES pipeline implemented in the MIDAS context. After careful inspection, we
decided that
the quality of the reduced spectra delivered by ESO
did not require a new reduction of the raw frames; therefore, we just
co-added spectra taken during different exposures, after applying
the correction for radial velocity shifts and normalization. 
Total exposure times range between
3600 -- 10800~s, depending on the star magnitude.
The final S/N ratios per resolution element are in the intervals
$\sim 20$ and 80 in the Be region and $\sim 120$ and 350 
in the Li region. 
The sample stars together with the log of the observations and achieved S/N
ratios are listed in Table \ref{sample}.
\section{Abundance and Error Analysis} \label{analysis}
Effective temperatures (\teff's) for the majority of M~67 stars were 
retrieved from Jones et al.
(\cite{jon99}, see also Randich et al. \cite{R06}) who, in turn, had 
derived them from dereddened B--V colors using 
the calibration of Soderblom et al. 
(\cite{sod93}). We determined \teff's for the two M~67
members not included in Jones et al. sample (S1034 and S1239) and
for NGC~2516 and Hyades stars
in the same fashion. For M~67 a reddening 
E(B--V)=0.04 was used, following Jones et al.,
while we assumed E(B--V)=0.01 and 0.12 for the Hyades and NGC~2516.
Adopted T$_{\rm eff}$ values are listed
in the second column of Table~\ref{results}.
A surface gravity $\log$~g=4.5 was assumed
for MS stars in NGC~2516 and the Hyades, while
for M~67 we used the spectroscopic values of Randich et al. (\cite{R06}).
Microturbulence was derived from $\log$~g
and T$_{\rm eff}$ based on the relationship
$\xi=3.2\times 10^{-4}(\rm T_{\rm eff}-6390)-1.3(log \rm g-4.16) + 1.7$~km/s
(Nissen \cite{nis81}).
\par Both Li and Be analyses were performed in the same way as in R02.
Li abundances were derived by means of equivalent widths (EWs)
of the Li~{\sc i} 6707.8~\AA~feature and using Kurucz's WIDTH9
(Kurucz 1993) code and model atmospheres. Our spectral resolution,
together with the low rotational velocity of all sample stars, allowed
us to resolve the Li line from the nearby Fe~{\sc i} 6707.4~\AA~feature
and hence there was no need to correct measured Li EWs for its contribution. 
EWs and derived Li abundances are listed in Cols.~3 and 4 of
Table~\ref{results}. Errors associated to our Li abundances
range between 0.08 and 0.17~dex. These combine the uncertainties in the EW
measurement (as reported in Col.~3 of the table, and which include also the
uncertainty due to the placement of the continuum) and in the stellar
parameters (i.e. how much the Li abundance is affected by changes in the
stellar parameters); as well known,
the dominant source is the high sensitivity of
Li to the effective temperature (0.08~dex per 100~K). 
Uncertainties in the stellar parameters were
taken from Randich et al (2006) for M~67 and are as follows:
$\pm70$~K in \teff, $\pm0.25$ in
$\log$~g, $\pm0.15$\,km/s in $\xi$. We assumed the same uncertainties
for the Hyades and NGC~2516.
For a more detailed
breakdown of all uncertainties affecting Li determination, we refer the
reader to Charbonnel \& Primas (\cite{cp05}).
\setcounter{table}{0}
\begin{table*}
\caption{Sample stars and log of the observations. Listed S/N ratios are
per resolution element.} \label{sample}
\begin{tabular}{lrlrrcc}
  & &  & & & & \\ \hline\hline
ID & V & (B--V)$_0$ & Obs. Date & Exp. time & S/N @ Be & S/N @ Li\\
 & &  &  & (sec)~~~~ & &  \\
  & &  & & & & \\ 
 \multicolumn{7}{c}{Main sequence stars}\\
  & &  & & & & \\
M~67 S990  & 13.43 & 0.525 &  Mar. 22,23 2002 & 7200 & 35 & 150\\ 
M~67 S995  & 12.76 & 0.521 &  Feb. 3 2002 & 5400 & 40 & 200 \\ 
M~67 S998  & 13.06 & 0.518 &  Jan. 7 2002 & 5400 & 35 & 175 \\ 
M~67 S2205 & 13.14 & 0.534 &  Mar. 1,4 2002 & 7200 & 40 & 200 \\ 
Hyades vB21  & 9.14 & 0.81 &  Dec. 3 2001  &  1200 & 75 & 350 \\ 
Hyades vB109 & 9.40 & 0.81 &  Dec. 23 2001 &  1800 & 80 & 350 \\ 
Hyades vB182 & 9.01 & 0.83 &  Dec. 21 2001 &  1200 & 80 & 350  \\ 
Hyades vB187 & 9.01 & 0.75 &  Nov. 29 2001 &  1200 & 80 & 350 \\ 
NGC~2516 CTIO-2 & 14.20 & 0.79 &  Dec. 19,23 2001 &  10800 & 20 & 150 \\ 
NGC~2516 CTIO-3 & 14.30 & 0.82 &  Dec. 20,22 2001 &  10800 & 20 & 150 \\ 
NGC~2516 CTIO-6 & 13.62 & 0.66 &  Dec. 30 2001    &   9600 & 30 & 180 \\ 
NGC~2516 DK213  & 14.01 & 0.76 &  Dec. 28 2001    &   5400 & 20 & 120 \\ 
  & &  & & & & \\ 
 \multicolumn{7}{c}{Blue stragglers and subgiants}\\
  & &  & & & & \\ 
M~67 S997  & 12.13 & 0.410 & Feb. 24 2002 & 3600 & 50 & 210 \\ 
M~67 S2204 & 12.89 & 0.417 & Mar. 8, Apr. 6 2002 & 5400 & 40 & 180 \\ 
M~67 S1034 & 12.65 & 0.567 & Apr. 15,16 2002 & 4800 & 35 & 200 \\ 
M~67 S1239 & 12.75 & 0.692 & Mar. 7,8 2002   & 5400 & 30 & 190\\ 
  & & & & & &  \\ \hline
  & &  & & & & \\ 
 \multispan{7}{Numbering for M~67 from Sanders (\cite{sand})\hfill}\\
 \multispan{7}{Numbering for the Hyades from van Bueren (\cite{vb})\hfill}\\
 \multispan{7}{Numbering for NGC~2516 from Jeffries et al. (\cite{jef98})\hfill}\\
\end{tabular}
\end{table*}
\setcounter{table}{1}
\begin{table*}
\caption{Effective temperatures, equivalent widths of the Li~{\sc i} 
6707.8~\AA~line,
and derived Li and Be abundances. These are in the usual
notation $\log$~n(X)=$\log$~N(X)/N(H)+12. Note that a positive offset
of $\sim 0.1-0.2$~dex in \nbe~might be present for stars cooler than 5400~K --
see text.} \label{results}
\begin{tabular}{lcccr}
   &  & &  &    \\ \hline\hline
star & \teff & EW(Li) & \nli & \nbe \\
 & (K) & (m\AA) &  &  \\
 &     &        &  &  \\ 
 \multicolumn{5}{c}{Main sequence stars}\\
 &   & &  &   \\ 
M~67 S990  & 6191 & $45\pm 4$ & 2.63$\pm 0.07$ &  0.96   $\pm 0.13$\\
M~67 S995  & 6210 & $17\pm 2$ & 2.10$\pm 0.08$ &  0.91   $\pm 0.13$\\
M~67 S998  & 6223 & $44\pm 2$ & 2.64$\pm 0.06$ &  0.91   $\pm 0.13$\\
M~67 S2205 & 6156 & $11\pm 3$ & 1.86$\pm 0.16$ &  0.81   $\pm 0.13$\\
Hyades vB21  & 5141 & $\leq 3$ & $\leq 0.27$   &  0.85   $\pm 0.21$ \\
Hyades vB109 & 5141 & $5\pm 1$ & 0.49$\pm 0.11$    & 0.80   $\pm 0.21$ \\
Hyades vB182 & 5079 & $10 \pm 1$ & 0.73$\pm 0.07$  & 0.80  $\pm 0.21$\\
Hyades vB187 & 5339 & $17 \pm 1$ & 1.27$\pm 0.06$  & 0.80   $\pm 0.21$\\
NGC~2516 CTI0-2 & 5238 & $147 \pm 3$ & 2.69$\pm 0.06$  & 1.16  $\pm 0.21$\\
NGC~2516 CTI0-3 & 5110 & $174 \pm 4$ & 2.80$\pm 0.07$  & 1.04  $\pm 0.21$\\
NGC~2516 CTIO-6 & 5659 & $124 \pm 5$ & 2.93$\pm 0.07$  & 1.21  $\pm 0.21$\\
NGC~2516 DK213  & 5305 &$108\pm 2$ & 2.41$\pm 0.06$    & 1.06  $\pm 0.21$\\
  &  & & &  \\ 
 \multicolumn{5}{c}{Blue stragglers and subgiants}\\
 &   & &  &   \\ 
M~67 S997  & 6700 & $\leq 5$  & $\leq 1.91$  &  0.31  $\pm 0.25$\\
M~67 S2204 & 6668 & $\leq 5$  & $\leq 1.89$  & $-0.20$  $\pm 0.29$\\
M~67 S1034 & 5969 & $\leq 3$  & $\leq 1.14$     & 0.12  $\pm 0.13$\\
M~67 S1239 & 5477 & $8\pm 1$  & 1.12$\pm 0.10$  & 0.11  $\pm 0.21$ \\
 &   & &  &   \\ 
 &   & &  &   \\ \hline
\end{tabular}
\end{table*}
\par
\begin{figure*}
\psfig{figure=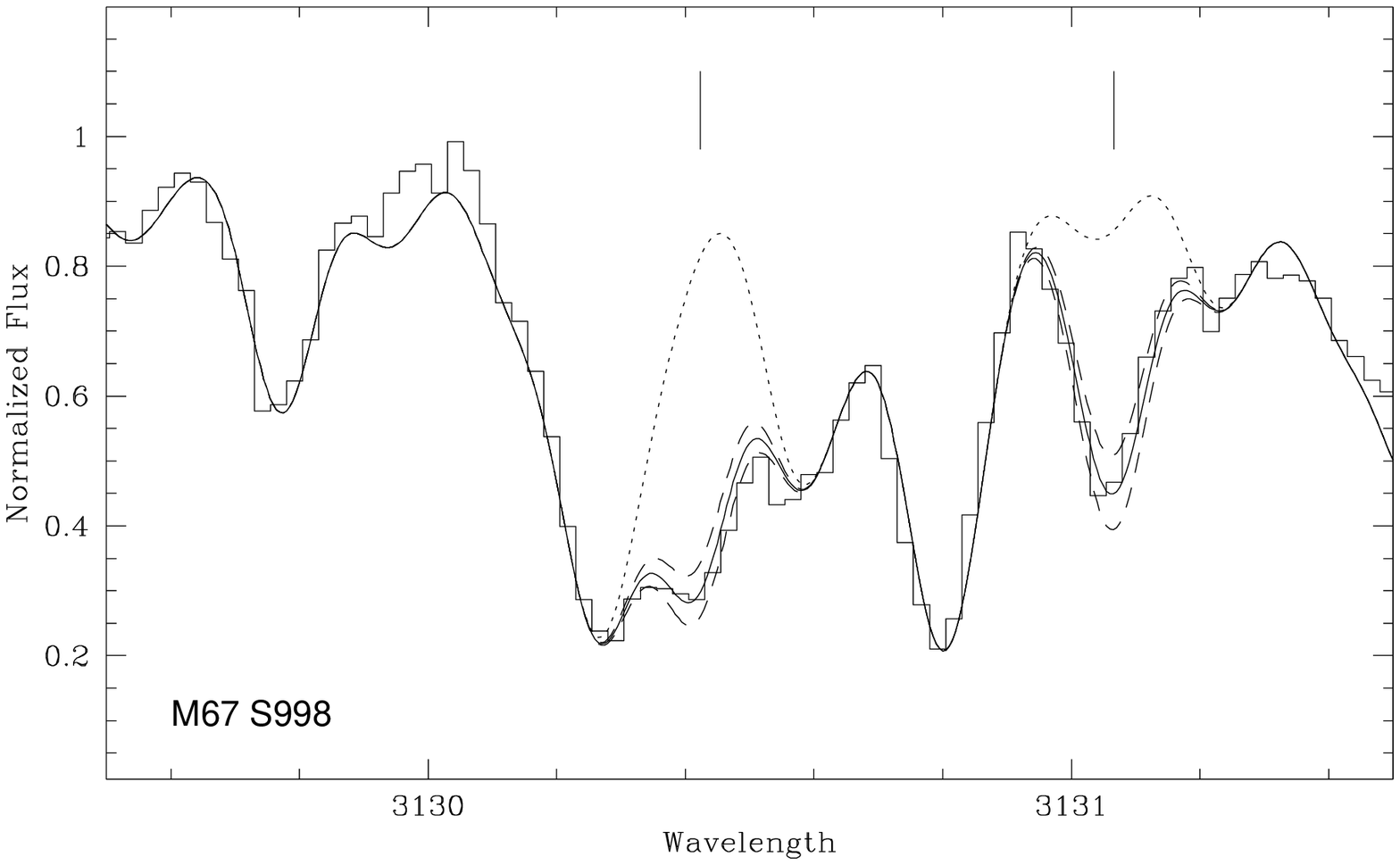, width=17cm}
\caption{The observed spectrum of the M~67 star S998 in the Be region with four
spectrum syntheses overplotted: our best-fit synthesis (continuous line), two
syntheses computed respectively with a variation of the Be content 
of $\pm0.15$~dex
(dashed lines) and one synthesis computed 
without beryllium (dotted line). The two vertical
bars identify the components of the \beii\,doublet.}
\label{figS998}
\end{figure*}

\begin{figure*}
\psfig{figure=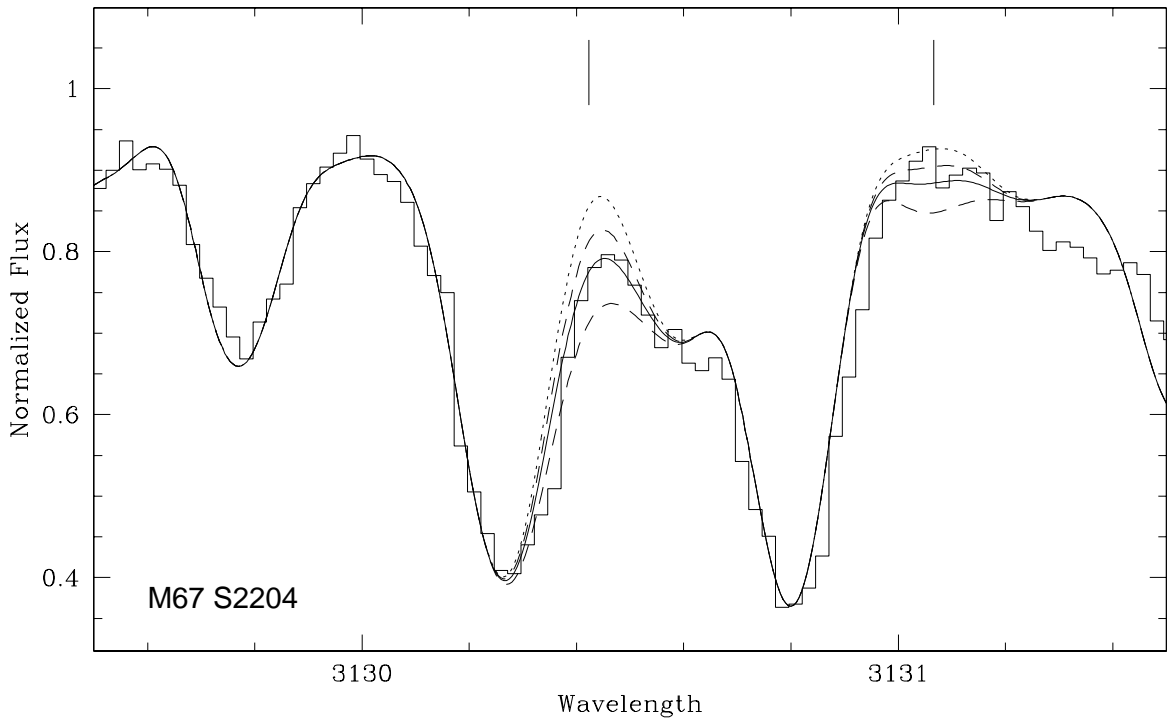, width=17cm}
\caption{Same as Fig.~\ref{figS998}, but for the M~67 star 
S2204 and for a variation in
the Be content of $\pm0.3$~dex.}
\label{figS2204}
\end{figure*}

\begin{figure*}
\psfig{figure=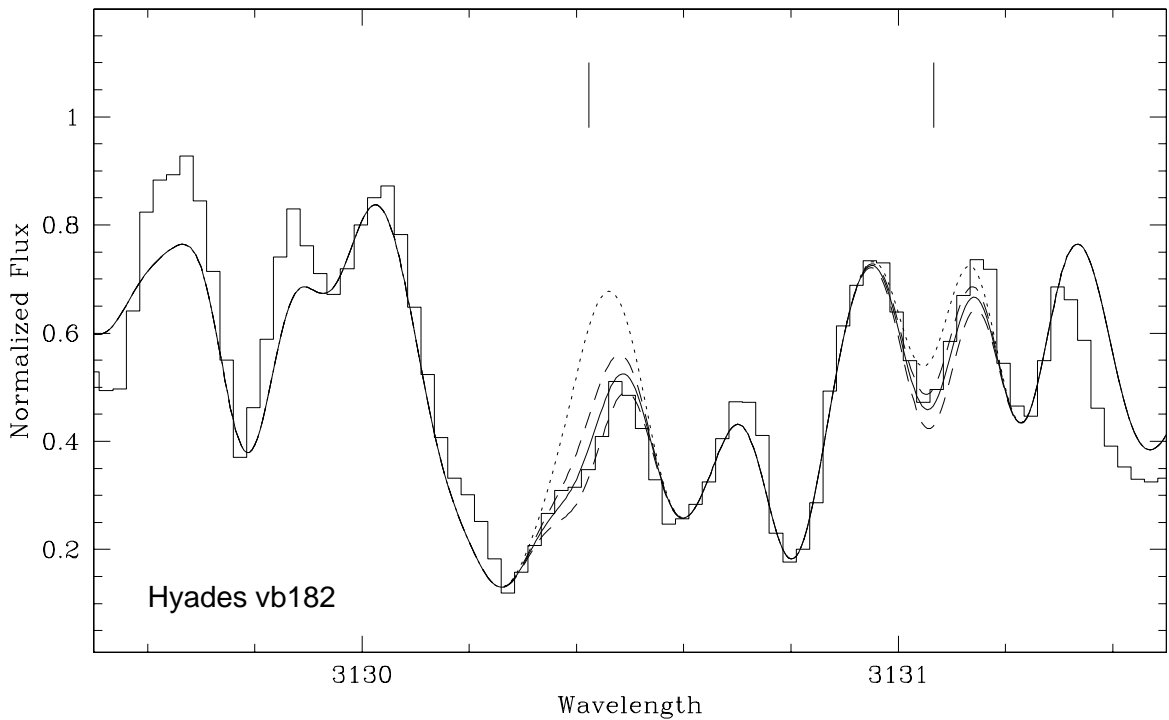, width=17cm}
\caption{Same as Fig.~\ref{figS998}, 
but for the Hyades star vB182 and for a variation
in the Be content of $\pm0.25$~dex.}
\label{figVB182}
\end{figure*}

\begin{figure*}
\psfig{figure=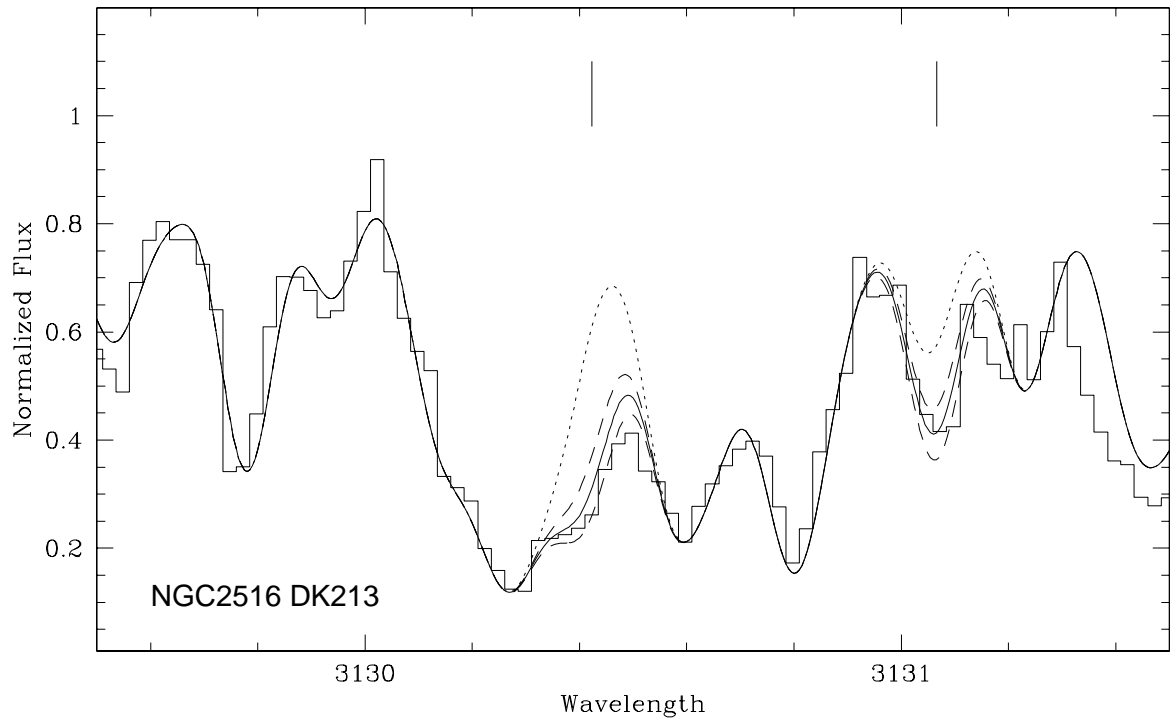, width=17cm}
\caption{Same as Fig.~\ref{figS998}, but for the NGC~2516 
star DK213 and for a variation
in the Be content of $\pm0.25$~dex.}
\label{figDK213}
\end{figure*}

 As mentioned,
Be abundances were determined using the same spectral synthesis
recipes as in R02 because we would like to combine both samples for our
final discussion. Namely, we synthesized a region of $\approx$5\AA,
around the \beii~doublet ($\lambda=3130.420$~\AA~and $\lambda=3131.064$~\AA),
with Kurucz's model atmospheres and SYNTHE code (Kurucz \cite{kur}) and with
the line list assembled and thoroughly tested by Primas et al. (\cite
{primas97}). Because of the severe blanketing present at these wavelengths,
and because we are working with solar (or over-solar) metallicity stars, the
normalization of the spectra is critical. Since no line-free region is
available around the \beii\, lines, we normalized our observed spectra by
dividing them for the highest number of counts in each spectrum, in a
15--20~\AA\,region centered on the \beii\,doublet; this step was performed
before multiple spectra for a given object were co-added. The spectra were
then further shifted (by a few \%, on average) to match the computed synthetic
spectra, but the overall continuum shape was maintained. A few examples of
spectrum syntheses are presented in Figs.~\ref{figS998}--\ref{figDK213}, where
we show our best-fit synthesis, two syntheses
computed by varying the Be content by a fixed amount, and one
synthesis computed without Be. 

Our final $\log$~n(Be) (as derived with 1D model atmospheres and including a
predicted Fe~I line in our list of transitions) and associated uncertainties
(see below) are listed in the last column of Table~\ref{results}. 

As a final remark, we note that Randich et al.~(\cite{R06}) derived
spectroscopic values of surface gravity for their sample stars in M~67
somewhat different from those assumed by R02; for this reason we carried
out a new analysis of the two stars (S988 and S994) whose new $\log$~g values
differed from the old ones by more than 0.25~dex (the assumed error in
$\log$~g). New and old abundances for these stars are listed in Table~\ref{new}.
In the following we will consider the new abundances derived in this work. 
\subsection{Errors in Be abundances}
\subsubsection{Random uncertainties}
\par In order to estimate the uncertainty associated to our Be abundances, we
computed the sensitivity of the derived Be abundances to changes of the
stellar parameters and of the placement of the continuum. Our final errors
do not include the uncertainty related to the atomic physics of the Be
transitions (the log~{\it gf} values are supposedly known with high accuracy)
or the absolute error on the model atmospheres, which is difficult to
properly estimate. Because our sample includes stars in different
evolutionary stages, we performed our sensitivity analysis for three stars
characterized by different stellar parameters: S998, vB182, and S2204. We
chose S998 as representative of our MS M~67 sample, vB182 as representative
of the cooler MS stars, and S2204 as one of the hottest stars in our sample
with very little Be left. 
Because the metallicity of these 
clusters is rather well determined, we assumed an uncertainty
of $\pm0.05$~dex on [Fe/H]. 
For the placement of the continuum, we estimated
that $\pm3\%$ is representative of this uncertainty. This number was derived
by normalizing our observed spectra with a different method, i.e. using the
computed synthetic spectra as the main reference to be matched. A summary
of our Be error analysis is provided in Table~\ref{taberrBe}, where the
numbers reported in the last row ($\sigma_{\rm tot}$) are the square root of
the sum in quadrature of all factors. We note that when the dependence
on some stellar parameters was not perfectly symmetric (when increasing
or decreasing that parameter by the chosen fixed amount), we always took
the average as the representative value. 
\setcounter{table}{2}
\begin{table}
\caption{New (present analysis) and old (R02)
beryllium abundances for S988 and S994 already analyzed
by R02.}\label{new}
\begin{tabular}{llcccc}\\ \hline
star & \teff & $\log$~g$_{\rm old}$ & \nbe$_{\rm old}$ & $\log$~g$_{\rm new}$ &  \nbe$_{\rm new}$\\
  &  & & & & \\
S988 & 6153 & 4.5 & 0.88 & 4.1 & 0.65 \\
S994 & 6151 & 4.5 & 1.16 & 4.1 & 0.86\\ \hline
\end{tabular}
\end{table}
\setcounter{table}{3}
\begin{table*}
\caption{Sensitivity of Be abundances to \teff, $\log$~g, 
$\xi$, [Fe/H], and placement of the continuum, for three stars of our
sample.}\label{taberrBe}
\begin{tabular}{cccc}
 & & & \\ \hline\hline
 & S998 & S2204 & vB182 \\ 
 & & & \\ 
\teff (K) & 6222 & 6668 & 5079 \\
$\pm70$~K & $\pm0.03$\,dex & $\pm0.07$\,dex & $\pm0.05$\,dex \\
$\log$~g & 4.0 & 4.0 & 4.5 \\
$\pm0.25$ & $\pm0.10$\,dex & $\pm0.08$\,dex & $\pm0.17$\,dex \\
$\xi$ (\kms) & 1.50 & 1.35 & 0.84 \\
$\pm 0.15$~\kms & 0.00 & 0.00 & $\pm0.05$\,dex \\
$\rm [Fe/H]$ (dex) & 0.00 & 0.00 & 0.13 \\
$\pm0.05$~dex & $\mp0.03$\,dex & $\mp0.10$\,dex & $\mp0.05$\,dex \\
$\pm3\%$ cont. & $\mp0.04$\,dex & $\mp0.25$\,dex & $\mp0.08$\,dex \\
\hline
$\sigma_{\rm tot}$ & $\pm0.13$\,dex & $\pm0.29$\,dex & $\pm0.21$\,dex \\
 & & & \\ \hline
\end{tabular}
\end{table*}
\subsubsection{Systematic uncertainties}
We remind from R02 that the list of lines we have used includes a predicted
line (i.e., not fully identified in the laboratory) of neutral iron at
$\lambda=3131.043$~\AA\, which is needed 
in order to obtain a satisfactory fit of the
$\lambda=3131.064$~\AA\,Be line in the solar spectrum. Our analysis of 
the Sun yields a photospheric solar Be abundance \nbe$_{\odot}=1.11$, in
excellent agreement with the value found by Chmielewski et al. (\cite{chm75}).
The atomic characteristics of this line were constrained by Primas et al.
(\cite {primas97}) on a large sample of stars, spanning a wide range of
stellar parameters (in temperature, gravity and metallicity). 
It is important to note, however, that Garc\'\i a L\'opez et al. 
(\cite{gar95}) proposed an
alternative solution to the unsatisfactory fit of the redder component of the
Be doublet, i.e. an increase of the oscillator strength of the Mn~I line at
$\lambda=3131.037$~\AA\,by $+$1.5~dex (from the value of $-1.725$ reported in
the Kurucz linelist). They found that such line becomes the dominant blending
transition of the redder component of the Be doublet for stars with effective
temperature below 5200~K, which prevented them from making a full detection of
Be in their lowest temperature stars. Because of this finding and of the fact
that, at solar (or above solar) metallicity,
the predicted line used by
Primas et al.~(\cite{primas97}) was tested only down to $\sim$ 5400~K 
(and our current sample reaches instead
5079~K), we decided to run a few more tests and to compare the effect of the two
proposed solutions, especially for the lowest temperature stars. 

First of all, we confirmed that both solutions do not have any important effect
for stars above 5400~K, except for allowing a better fit of the blue wing of
the $\lambda=3131.066$~\AA. As we tested lower effective temperatures, we
found that the two solutions differ by 0.2~dex in the final derived Be
abundance in the 5000--5100~K range, while in the 5250--5400~K interval
the difference
amounts to 0.15--0.10~dex, with our solution always yielding
lower Be abundances. In other words, for stars below 5400~K the 
strength of the predicted
Fe~I line is larger than that of the the Mn~I and the difference
increases with decreasing temperature. Note, however, that our
proposed solution still 
allows us to measure Be abundances down to the lowest \teff~in our sample.
Whereas there is no solid argument in favour of one solution with
respect to the other, depending on the solution one
decides to follow, the lowest temperature stars may end up having slightly
different abundances, with the offset depending on the temperature. 
For this reason we are not able to derive
a definitive {\it absolute} Be abundance and thus to quantify the 
exact amount of Be depletion suffered by Hyades members in our sample;
nevertheless the {\it relative} abundances of NGC~2516 and Hyades stars with
similar temperatures suggest that the latter have undergone some depletion.

\section{Results} \label{sect_results}
\subsection{Comparison with previous studies}
Our Li abundances for most stars
are in very good agreement with the values derived in previous studies
(Jeffries et al. \cite{jef98} for NGC~2516; Thorburn et al.
\cite{thorb} for the Hyades; Jones et al.~\cite{jon99} and
Balachandran~\cite{bala95} for M~67). For one star only, 
CTIO-6 in NGC~2516, we derive an abundance somewhat lower than 
(but given the errors still consistent with) the value 
of Jeffries et al. (\cite{jef98}); namely, we obtain \nli=2.93 to
be compared with value \nli=3.04. Note however that Jeffries et al.
(\cite{jef98}) measured a much higher EW (212~m\AA),
which in our abundance scale would correspond to \nli~$\sim 3.4$.
Also, we detected the Li line in the spectra
of two stars for which only upper limits were available in the
literature 
(S2205: \nli$_{\rm pres.}$=1.86 vs. \nli$_{\rm lit.}$ $\leq 2.34$ 
--Jones et al. \cite{jon99};
S1239: \nli$_{\rm pres.}$=1.12
vs. \nli$_{\rm lit.}\leq 1.1$ --Balachandran \cite{bala95}) and 
were able to put
more stringent upper limits on Li abundances for three stars: namely 
S997 and S2204 (\nli$_{\rm pres.}\leq 1.91$ vs. \nli$_{\rm lit.}\leq 2.46$; 
\nli$_{\rm pres.}\leq 1.89$
vs. \nli$_{\rm lit.}\leq 2.36$ --Shetrone \& Sandquist \cite{shs00}) and
S1034 (\nli$_{\rm pres.}\leq 1.14$ vs. \nli$_{\rm lit.}\leq 1.6$ 
--Balachandran \cite{bala95}). 

Our study provides the first determination of Be for
all the sample stars, but vB21; this star was included in the study of
Garc\'\i a L\'opez et al. (\cite{gar95}) who measured \nbe=0.90$\pm 0.25$,
in agreement with our own determination (\nbe$=0.85\pm 0.21$).
\subsection{Be vs. \teff}
In Fig.~\ref{abteff} we plot \nbe~(upper panel) and \nli~(lower panel)
as a function of  T$_{\rm eff}$ for our
sample stars and the sample of R02. 
The present M~67 sample
includes four new stars warmer than 6150~K, but no cooler cluster members;
therefore the conclusions of R02 for solar-type stars remain unchanged:
independently of their absolute Be abundance, 
almost all the stars in the range $5600 \leq$~\teff~$\leq 6150$~K
show, within the errors, the same relative abundance, in
spite of the different levels of Li depletion (see lower panel of Fig.~
\ref{abteff}). On the other hand, 
stars warmer than 6150~K both in the present and
in the R02 samples have lower abundances, suggesting that
they have suffered a certain amount of Be depletion.
Note that, as discussed in detail in Sect.~\ref{disc_in} below, we consider
the abundance of SHJM2, a member of the 55~Myr old cluster
\object{IC~2391}, as representative of the value of the initial Be abundance
in our scale.

As to stars cooler than 5400~K --note that our sample does not include stars
with \teff~in the range 5400--5600~K--, Be abundances of
NGC~2516 members are consistent with no depletion
in spite of the factor $\sim$2--8 Li depletion. 
The significantly more Li-poor Hyades have lower Be abundances:
the average value of the four Hyades stars
is $\sim 0.25$~dex (or a factor 1.8) below the average of the
three NGC~2516 stars cooler than 5400~K. This result would not change
if the abundances were derived assuming the solution proposed
by Garc\'\i a L\'opez et al. (\cite{gar95} --see discussion in 
Sect.~\ref{analysis}). 

Finally, both the two blue stragglers (S997 and S2204) and the
two evolved M~67 members are characterized by significant amounts of
Li and Be depletion.
\section{Discussion}
\subsection{Initial beryllium abundance}\label{disc_in}
As discussed by R02, the Be abundance of the young star SHJM2 in 
IC~2391 (\nbe=1.11), as well 
as our value for the photospheric solar Be (\nbe=1.11),
are about a factor of two
lower than the meteoritic abundance (\nbe~$=1.42$ --Anders \& Grevesse
\cite{ag89} \footnote{
A slightly lower value, \nbe~$=1.38\pm0.08$, has been recently reported
by Asplund et al.~(\cite{asplund05}).}). Abundances for stars in the
present study and in R02 are all below this value. Also,
maximum Be abundances similar to ours are
found by  Garc\'\i a L\'opez et al. (\cite{gar95}) for the Hyades
and by Santos et
al. (\cite{santos}) for their solar metallicity stars;
both studies derive a Be abundance for the Sun very close to
our determination. 
IC~2391 has a solar metallicity (Randich et al. \cite{R01}) and thus
its initial Be abundance is most likely the same as for the solar system.
The difference between our \nbe~values for the Sun and SHJM2
and the meteoritic Be abundance suggests that either
{\bf a)} both the Sun and
SHJM2 have depleted a factor of two Be; or
{\bf b)} they have not undergone any depletion, the difference with
the meteoritic value is due to
systematic effects in the adopted analysis, and the \nbe~value of
SHJM2 is indicative of the initial Be abundance in our scale.
We mention in passing that the discrepancy between the solar photospheric Be 
and meteoritic abundance has been known since long time
and it has been discussed in several works. 

\begin{figure*}
\vspace{-5cm}
\psfig{figure=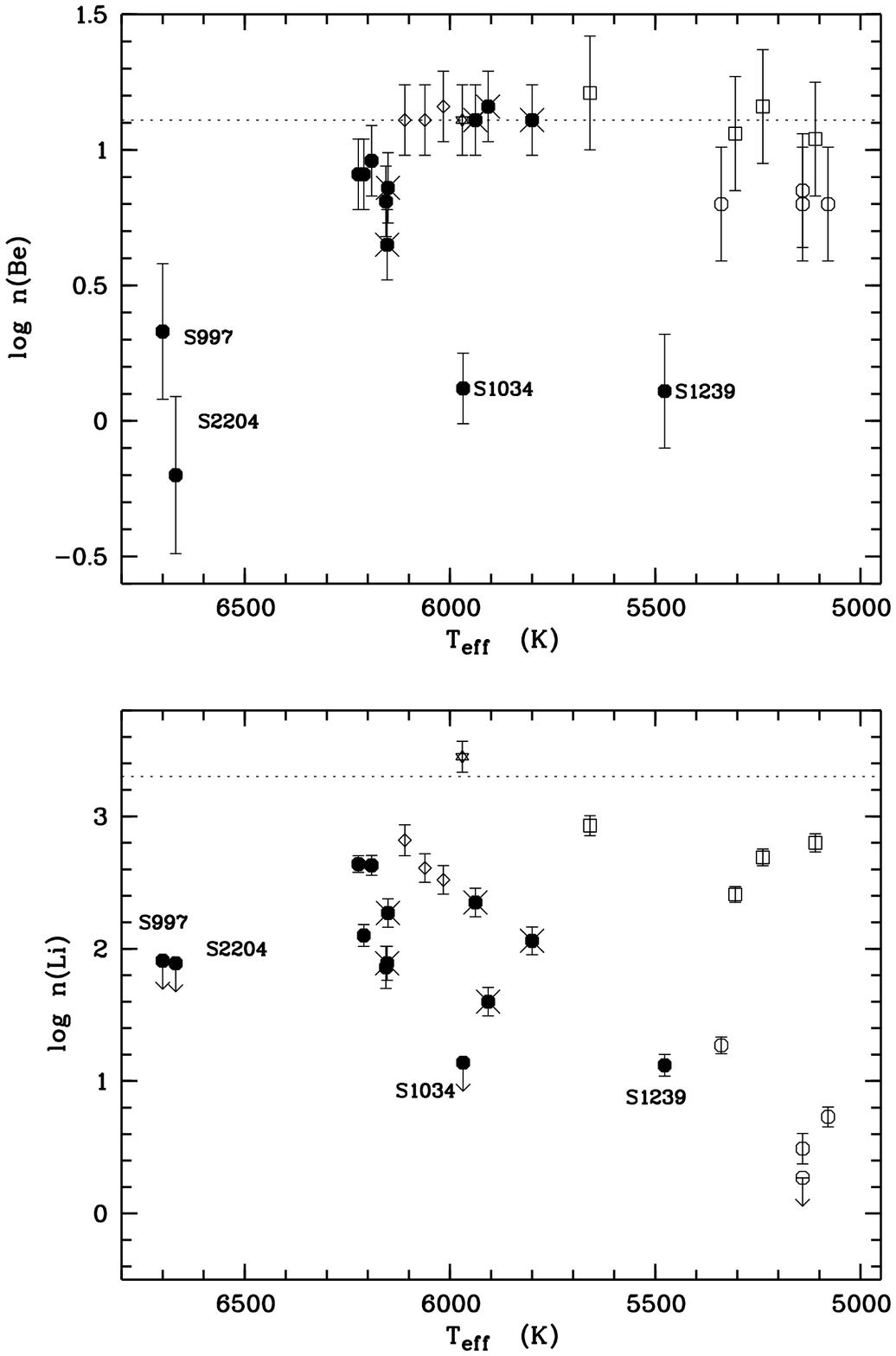, width=17cm}
\caption{Upper panel: Beryllium abundances (\nbe, in the usual
logarithmic scale where $\log$~n(H)=12) as a function effective temperature
(\teff) for our sample stars plus the sample of R02.
Symbols are as follows: M~67: filled circles (present
sample) and crossed filled circles (R02); NGC ~2516: open squares;
Hyades: open circles; IC~4651 (R02): open diamonds; IC~2391 (R02): open
star symbol. 
Lower panel: same as upper panel, but Li --\nli-- abundances are plotted.
Names of M~67 stars out of the MS are indicated in both panels. Horizontal lines
denote initial Be and Li abundances in our scales.
} \label{abteff}
\end{figure*}

\begin{figure*}
\vspace{-4cm}
\psfig{figure=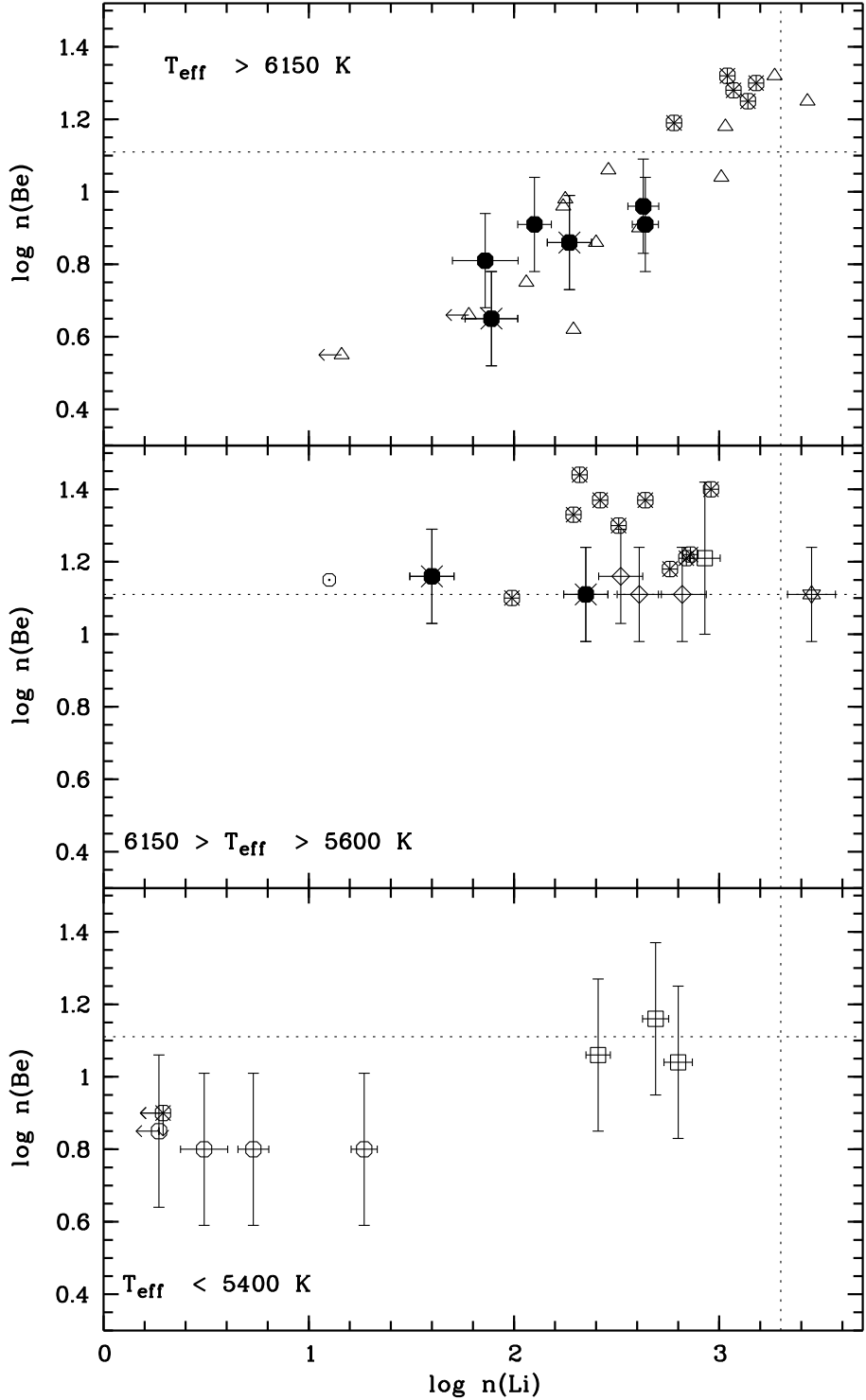, width=17cm}
\caption{\nbe~as a function of \nli~for our sample stars, the sample of R02,
and the Hyades from Boesgaard et al.~(\cite{bk04}) and Garc\'\i a
L\'opez et al.~(\cite{gar95}). 
Upper panel: stars warmer than 6150~K. Middle panel:
stars with 5600 $\leq$~\teff~$< 6150$~K; Lower panel: stars cooler than
5400 K.
M~67 members out of the MS are not plotted in the figure.
The Sun is also shown in the middle panel. 
Symbols are the same 
as in Fig.~\ref{abteff}; Hyades from the literature, not plotted
in Fig~\ref{abteff}, are represented here as open circles with stars (in all
panels for \teff $\leq 6300$~K) and open triangles (upper panel, stars
with \teff $\> 6300$~K).
Horizontal and vertical dashed lines denote
initial Be and Li abundances in our scales.} \label{nlinbe}
\end{figure*}

\begin{figure*}
\psfig{figure=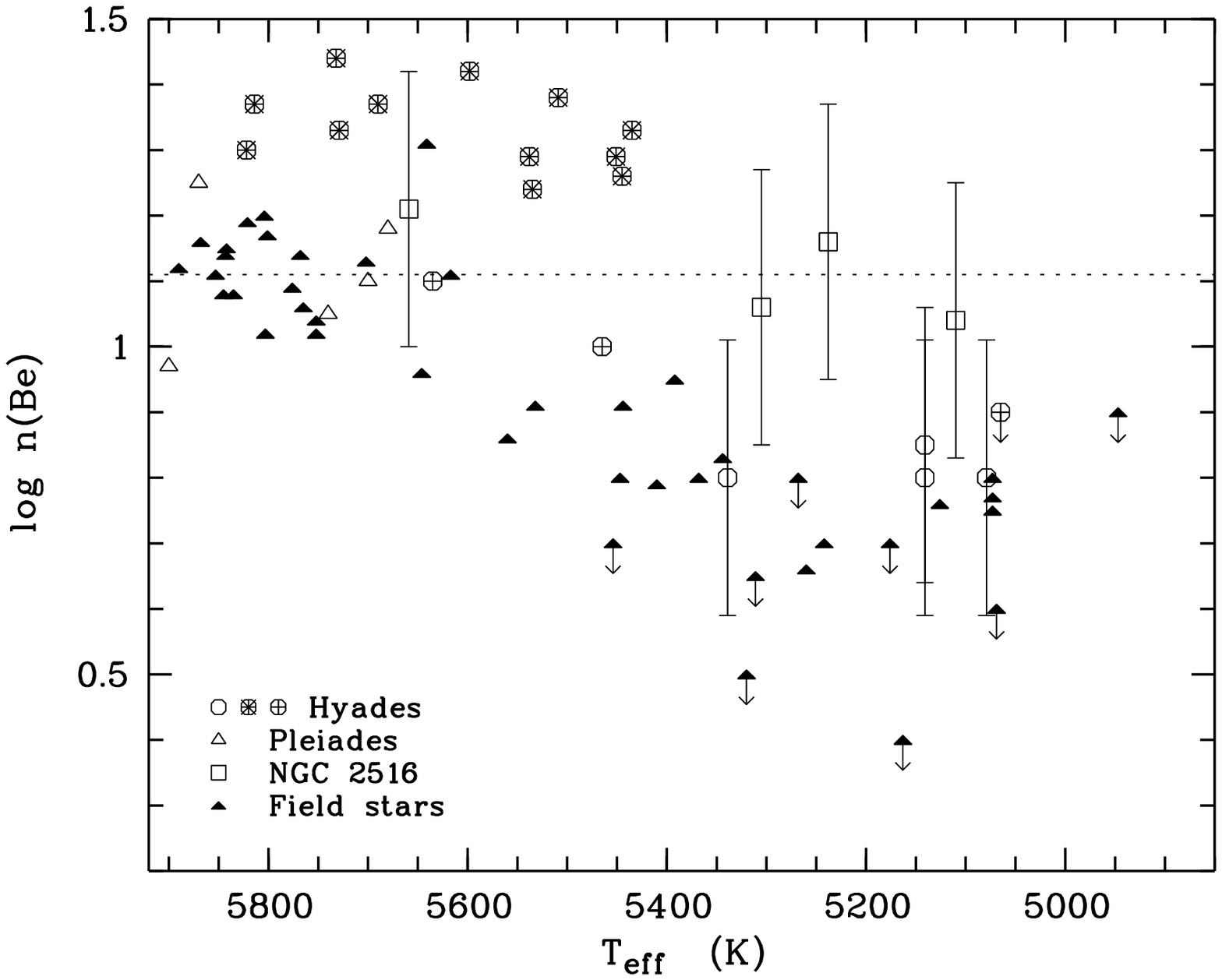, width=17cm, angle=-90}
\caption{\nbe~as a function of \teff~for NGC~2516 (150~Myr -open
squares), the Hyades (600~Myr --present sample: open circles;
sample of Boesgaard et al.~\cite{bk04}: open circles with stars; sample
of Garc\'\i a L\'opez et al.~\cite{gar95}: open circles with plus symbols),
Pleiades (125~Myr --open triangles; data of Boesgaard et al. \cite{bk03}),
and field stars ($>1$~Gyr --filled
triangles; sample of Santos et al.~\cite{santos}).
The horizontal line denotes the initial Be abundance in our scale.
}
\label{age}
\end{figure*}

Hypothesis {\bf a)} and, specifically, the fact that
SHJM2 has depleted some Be, would imply that Be burning occurs
during the pre-main sequence (PMS) phases, at variance with
the observational evidence of no PMS Li depletion for stars in
this \teff~range (Randich et al. \cite{R01} and references
therein);
it would indeed be hard to explain how this star depleted some Be
while maintaining its initial Li.
For this reason, we regard possibility {\bf a)}
as very unlikely and favor hypothesis {\bf b)} above. At the same time,
since the metallicities of M~67 and NGC~2516 are also solar (Randich et al.
~\cite{R06}; Jeffries et al.~\cite{jef98}), it is unlikely that these clusters 
had different initial Be abundances due to Galactic
evolutionary effects. We conclude therefore that all the sample
stars with \nbe~consistent with the abundance of SHJM2, including the Sun,
have most likely not undergone any Be depletion.
A strong support to this hypothesis is given by the 
analysis of Balachandran \& Bell ({\cite{bb98}) of the solar Be who 
showed that, taking into account an additional
amount of UV opacity, needed to solve the inconsistency
between solar oxygen abundance derived from IR and UV lines,
accounts perfectly for the missing Be and makes the discrepancy between the
solar photospheric Be and the meteoritic value disappear. The
conclusion that the photospheric solar Be is consistent with 
the meteoritic one was also reached by  Asplund et al. (\cite{asplund04});
our assumption hence appears fully justified. We finally mention
that a higher value of the solar Be abundances was derived by
Boesgaard et al. 
(\cite{bk03}: \nbe$_\odot=1.30$), by using 2002 version of MOOG
which includes all of the Kurucz opacity edge;
also, for undepleted cluster
stars in their sample, they derive maximum abundances higher than ours
(and in particular than that of SHJM2). These differences suggest
the presence of
an offset between the abundance scales.
\subsection{Be and Li depletion}\label{MS}
In Fig.~\ref{nlinbe} we plot \nbe~as a function of \nli~for our sample
stars, the sample of R02, and the Hyades (from Boesgaard
et al.~\cite{bk04} and Garc\'\i a L\'opez et al.~\cite
{gar95}).
In the upper, middle, and lower
panels we show stars in three temperature ranges: namely,
6300$>$\teff$\geq 6150$~K, 5600~$\leq$\teff$< 6150$~K, 
and \teff$< 5400$~K.
This subdivision broadly corresponds to the different regimes of Be depletion
seen in Fig.~\ref{abteff}; specifically, stars that show both Li
and Be depletion, stars that show Li depletion, but no Be depletion,
stars that show Li depletion and may show Be depletion.
Whereas the precise boundaries
between these three intervals might be slightly different, it is important
to subdivide the sample stars in subgroups since, as we discuss below,
they likely correspond to different dominant mixing processes. 
In the upper panel we show for comparison also Hyades stars
on the red side of the Li gap. We finally note that
our sample does not include stars
with \teff~between 5400 and 5600~ĸ; this \teff~range is thus not represented
in the figure.

The figure evidences distinct behaviors for stars in the three
subsamples. Namely,
the middle panel, which includes only stars from R02
plus the Hyades,  clearly
indicates the absence of a Be vs. Li correlation for solar-type
stars in the 6150--5600~K range in both old and young clusters: they span
two orders of magnitude in Li abundances, they have different ages and
metallicities, but they share the same Be content.
This in turn implies that the mixing mechanism responsible for MS
Li depletion in this temperature range does not extend deep enough
to cause also Be depletion.
On the other hand, the enlarged sample with respect to R02 shows
that a correlation between Li and Be abundances is present
for warmer stars in M~67; these stars rather well fit into
the Be vs. Li trend for the Hyades on the red side
of the Li gap. 

Finally, the bottom panel and, in particular the difference between Be
abundances of the Hyades and NGC~2516 members, suggest that in
the Hyades cool stars Li depletion is 
accompanied by some amount of Be depletion. It would be not
formally correct to claim a Be-Li correlation for these stars, since
NGC~2516 members cover a large range in Li abundance,
but have not depleted any Be; also, Hyades stars themselves
show different amounts of Li depletion, but the same Be content.
Nevertheless, the figure shows that, in this \teff~regime,
at variance
with solar-type stars, but similarly to warmer objects (\teff$> 6150$~K),
the mechanism responsible for Li depletion can reach deep enough to
cause also Be depletion. We stress again that, whereas the absolute Be 
abundances of the coolest sample stars in the figure would be higher had
we used the solution of Garc\'\i a L\'opez et al.~(\cite{gar95}),
the relative depletion would remain the same.
\subsection{Cool stars: Be as a function of age}
Garc\'\i a L\'opez et al. (\cite{gar95}) already
found evidence of Be depletion for two cool Hyades members, while
Santos et al.~(\cite{santos}) detected Be depletion in several old
field stars cooler than about 5600~K. 
Our results not only reinforce the conclusion that
cool stars might deplete Be, but most importantly,
the comparison between stars with similar temperatures and different
ages allows us to put constraints on the age when Be depletion starts on the MS.

In Fig.~\ref{age} we plot \nbe~as a function of \teff~for NGC~2516, the
Hyades (present, Garc\'\i a L\'opez et al. \cite{gar95} and Boesgaard
et al.~\cite{bk04} samples), the Pleiades from Boesgaard et al. (\cite{bk03}),
and old field stars from Santos et al. (\cite{santos}). 
Note that the original sample of Santos et al. is mostly composed by stars 
older than 1~Gyr and we do not
plot in the figure the four young stars in their sample.
The figure shows that Be depletion starts being present below $\sim 5600$~K:
for \teff~between 5400
and 5600~K Be depletion is seen only among old field stars,
while for temperatures below
5400~K, Be depletion is seen also among the Hyades, but
not among NGC~2516 members. This implies that the age at which Be
depletion begins in cool stars increases with increasing \teff: 
namely, stars cooler than $\sim 5400$~K 
start depleting Be between 150 and 600~Myr, depletion starts
after 600~Myr for stars in the 5400--5600 interval, while stars warmer
than 5600~K (but cooler than 6150~K), including the Sun, never deplete Be.

Li depletion in MS stars cooler than 5500--5400~K is normally explained
with convection, since the convective envelope reaches deep
enough to mix Li-poor material into the surface.
The question then arises whether Be depletion
is also expected for these stars
based on standard models. In order to address this
question, 
we calculated evolutionary tracks for a 0.9 $M_{\odot}$ star, 
corresponding to the mass of our coolest stars in the
Hyades (see below), using the models by Sestito et al. (\cite{sesti_06}).
These were computed employing an updated
version of the evolutionary code FRANEC (e.g. Chieffi \& Straniero
\cite{chieffi}) and adopting updated physical inputs (equation of
state, opacity tables, nuclear cross sections, etc., see Sestito 
et al.~\cite{sesti_06} for further details). 
As customary in standard models, the only mixing mechanism included is
convection, whose extension was determined by adopting the classical
Schwarzschild criterion and the mixing length formalism.
Element diffusion has been included during the MS 
(Ciacio et al.~\cite{ciacio97}, 
with diffusion coefficients from Thoul et al.
\cite{thoul}).
Three different models were computed; namely, 
we adopted
two different solar mixtures: the solar composition by Asplund et
al. (\cite{asplund04}; hereafter As04) 
and that by Grevesse \& Noels (\cite{grevesse93}; hereafter GN93).
The first track computed is the ``standard'' model by Sestito et
al. (\cite{sesti_06}; model 1 in their Table 1): As04 mixture, $Z=0.013$
(corresponding to [Fe/H]=0 for this composition), He content $Y=0.27$, and
mixing length parameter $\alpha=1.9$.
The other two models have the same He abundance and $\alpha$ but the Fe content
of the Hyades
([Fe/H]$=+0.13$) has been considered, in one case adopting the As04
mixture (and the resulting global metallicity is $Z=0.018$), while in
the other using the GN93 composition ($Z=0.025$).
We mention that at the age of the Hyades, the [As04, $Z=0.013$] model has
an effective temperature of 5400 K, while
the [As04, $Z=0.018$] model has \teff=5200~K
and the [GN93, $Z=0.025$] one has \teff=5130~K,
consistent with the temperatures of Hyades stars in our sample.

In Fig.~\ref{tcz} we plot the temperature at
the base of the convective envelope ($\rm T_{\rm{CZ}}$) predicted by
our models as a function of time: the figure clearly shows that, for 
all the models, $\rm T_{\rm{CZ}}$ is slightly higher
than the Be ignition temperature only for ages younger than about 20 Myr,
implying that Be depletion can in principle occur only during PMS, while
no Be depletion is expected at older ages.
This is shown in Fig.~\ref{nbe_age} where we plot 
the resulting evolution of Be/Be$_0$ as a function
of age: consistently with the discussion above, none of the three models 
predict significant Be depletion. The model with solar metallicity
predicts a slightly lower amount of Be depletion with respect to
the other two, as expected (the higher is $Z$, the higher is the
efficiency of convection). However, and most importantly, 
the difference between the solar metallicity track and the two models
with the Fe content of the Hyades is almost negligible, and in all
cases a very small amount of Be is depleted with respect to
the initial value.
In particular, we note that even PMS Be depletion is always very small (almost
negligible), since
$\rm T_{\rm{CZ}}$ reaches values only slightly higher
than the temperature of Be ignition and, furthermore, the exact value of
ignition temperature depends on the physical assumptions in the code.
We mention in passing that this result is in agreement with the early
calculations of Bodenheimer~(\cite{bod66}.)

Our conclusion is that
Be depletion observed in the cool Hyades with respect to NGC~2516
cannot be explained
only by convective mixing and cannot be reproduced in the framework of standard
models. An additional, or extra-mixing
mechanism must be present, analogously to what invoked to explain Li
depletion in more massive, warmer stars. Most obviously, the
mechanism is not necessarily the same. Our Be determinations suggest
that this physical process starts being efficient
after an age of $\sim$150 Myr.
\subsection{Stars out of the main sequence}
\subsubsection{Subgiants}
M~67 subgiants S1034 and S1239 studied here were also included in the
Li survey of Balachandran (\cite{bala95}),
who derived for the two stars masses and ZAMS temperatures equal to 
1.282~M$_{\odot}$ and 6349~K (S1034) and 1.3~M$_{\odot}$ and 6450~K (S1239);
as argued
by Balachandran, their low Li abundances strongly support the hypothesis
that they have depleted Li during the MS phases and have evolved 
from the Li gap region. Our study
provides the first measurements of Be in stars evolved from the gap,
thus allowing us
to carry out further comparison with model predictions.
Both stars show similar factors of Be depletion (a factor of about 10 --see
Table~\ref{results}), suggesting the presence of a Be gap,
similarly to what found for the Hyades MS stars
Boesgaard \& King~(\cite{bk02}).
As discussed at length in the series of papers by Boesgaard and
collaborators and mentioned in Sect.~\ref{intro}
the tight correlation between Be and Li depletion
for stars in the gap suggests that 
light element depletion is driven by rotational
mixing.
Our Be determinations for the two subgiants are consistent
with this hypothesis.

More quantitatively, 
Sills \& Deliyannis (\cite{sd00}) presented evolutionary models of
subgiants in M~67 including three mixing processes (diffusion, mass
loss, rotational mixing) and following the evolution of both
Li and Be abundances. They concluded that Li measurements are in good
agreement with models including mixing driven by rotation;
they stressed however that Be data for subgiants in M~67 would 
also be very helpful in order to constrain the various free parameters
entering the models.
Our measurements of Be for two subgiants in M~67 represent a first step towards
that direction. In Fig.~\ref{sills} we compare our results with the predictions
of the model of Sills \& Deliyannis including rotationally induced mixing; 
while the other models discussed in that paper are not shown, since
they were already ruled out the basis of Li alone.
Note that the \teff~sequence in the figure represents an
evolutionary sequence for a given mass (1.26~M$_{\odot}$ -somewhat below
the value given by Balachandran~\cite{bala95} for subgiants in M~67) rather than
a sequence of different masses. 
The figure evidences a very good 
agreement between the model and our measurement for S1239, the cooler,
slightly more massive and more evolved star, whose Be content is the result
of MS depletion plus post-MS dilution. 
The agreement is considerably worse for the less evolved S1034, 
whose Be content
is the result of MS depletion only, and which
appears more Be depleted than predicted (but also
its Li abundance is below model predictions --see Fig.~4 in
Sills \& Deliyannis). We suggest that the reason
for this discrepancy could be an initial higher rotational velocity for this
star than the one assumed by Sills \& Deliyannis in their model
(v$_{\rm in.}=10$~km/s), which would result in a larger amount of
light element depletion on the MS.
\subsubsection{Blue stragglers}
Light elements also provide an useful tool
to investigate the mechanism of formation of blue stragglers and, in particular,
whether they are the product of collision processes or binary mergers.
To our knowledge, no quantitative models exist to-date presenting predictions
for Li and Be abundances for the different scenarii of blue straggler
formation. However, Sills et al. (\cite{sills97}) qualitatively
suggested that in the case 
a blue straggler is the result of a binary merger, it should not have
observable Li; on the other hand, collisions would not result in
any Li (and a fortiori Be) destruction.

The first observations of light elements in blue stragglers 
belonging to M~67
were performed by Pritchet and Glaspey~(\cite{pg91}), who
measured Li in seven stars; they did not detect
the Li line in any of them and concluded that
some form of mixing, possibly due to binary coalescence, had affected
the outer envelope of blue stragglers. We note however, that upper
limits for most of their sample stars were indeed consistent with
abundances of Li-poor MS cluster stars.

In a more recent study, Shetrone \& Sandquist (\cite{shs00}) 
carried out new Li measurements of blue stragglers in M~67, including
both S997 and S2204, the two stars studied here.
We refer the reader
to Shetrone \& Sandquist (\cite{shs00}) for a detailed discussion of 
the properties of the two stars; we just recall that the high eccentricity
binary nature of one of the two (S997) suggests that it is unlikely a binary
merger, which is instead the most plausible scenario for the
formation of S2204. Shetrone \& Sandquist (\cite{shs00}) did not detect the Li 
line in either star; although the non-detection of Li 
might favour the merger scenario,
these authors point out that their upper limits 
eventually do not provide any clues on the formation mechanism. 
The upper limits are not very stringent and are 
consistent with the abundances of MS unevolved cluster stars (i.e., the
presence of Li cannot be ruled out); furthermore, even a low Li
abundance could be explained by the fact that the two stars fall in the
Li gap \teff~interval and thus could have undergone significant depletion
after the blue straggler is formed in the collision scenario. 

Until the present study no Be determinations for blue stragglers
in open clusters were
available in the literature. We have now determined Be abundance for S997 and
S2204, finding that they are both depleted, but affected by different 
amounts of Be depletion;
namely, a factor of about 6 and 20 for S997 and S2204, respectively.
At variance with Li, the low Be abundances are not consistent
with those of MS stars in M~67; 
as in the case of Li, we cannot completely exclude that the two stars have
depleted Be after formation. Nevertheless, we 
notice that Be abundance for star S2204
is much below the values measured for stars in the Be gap. Thus, whereas 
Be abundance in S997 would not be inconsistent with either 
scenario, we suggest that the factor of three higher depletion measured
in S2204 might be an indication of 
a different formation of the two systems, 
with this star being the product of a binary merger.
\section{Conclusions}
We have carried out Be observations in three
open clusters of different ages: NGC~2516, the Hyades, and M~67.
The present study, together with available data from R02 for M~67
and from other previous Be surveys of open clusters,
allows us to cover a grid of MS stars which span from 
150 Myr to 4.5 Gyr in age and from the Li gap down to 5100 K
in effective temperature. 

We confirm that stars in different \teff~interval are characterized
by different amounts of Be depletion and Be vs. Li behaviors; 
namely, {\bf i.}
stars warmer than 6150~K are characterized by
some Be depletion and show a Be vs. Li correlation; 
{\bf ii.} Li and Be are instead not correlated for solar-type stars.
Whereas Li depletion ranges between null (for the young star in IC~2391)
and a factor larger than 100 (in the Sun and the Li poor stars
in M~67), none of the stars show significant Be depletion;
{\bf iii.} the Hyades stars cooler
than 5400~K are a factor of $\sim 1.8$ more Li depleted
than stars of similar temperature
in the younger NGC~2516;} {\bf iv.} subgiants and blue stragglers
are severely depleted in Be.

The result for warm stars is consistent with the extensive work 
carried out by Boesgaard and collaborators and, in particular, with
the observed Li-Be correlation for stars in the Li gap in the Hyades
and other clusters of similar age.
As mentioned in Sect.~\ref{intro}, the presence and slope
of this correlation are in agreement with the predictions of models including
slow mixing driven by rotation. Be depletion in the two subgiants in M~67
that are evolved from stars in the gap is also consistent with
rotationally induced mixing and our measurements actually can provide
additional constraints on theoretical models.
On the contrary, the lack
of a correlation between Li and Be for solar-type stars implies that
the main process at work in these stars must be different.
R02 indeed ruled out models including mixing
induced by rotation and diffusion which predict a Be-Li correlation.
They noticed that models including gravity waves would well fit
the observed Be vs. Li distribution, but those models would
not be able to explain the observed scatter in Li in M~67.

In a more realistic approach one might imagine that
various processes (rotationally driven mixing, 
meridional circulation, diffusion, gravity waves, tachocline) 
are all present at the same time; different mechanisms
might then either become dominant or counter-balance in different 
temperature regimes. In this context,
very recently Charbonnel \& Talon (\cite{ct05}) presented a model including
both mixing driven by rotation and gravity waves. They showed that
this model is able to reproduce the solar Li abundance as well as the
internal rotational profile of the Sun and that models with different
initial rotational velocities would result in a range of Li abundances,
consistent with the observed scatter in M~67. 
Since Charbonnel \& Talon~(\cite{ct05}) did not provide any 
predictions for the evolution of Be abundances, we cannot
make any quantitative comparison with the results of the present study.
We also mention in passing
that their model would not be able to reproduce the plateau
in Li abundances that is observed for ages greater than $\sim 2$~Gyr
(Randich \cite{r_cast} and references therein). More work is therefore
needed on theoretical grounds; we stress however that considering
the interplay between different processes might indeed represent the way
towards the solution to the light element depletion problem.

As to cooler stars, whereas between 5600~K and 5400~K Be depletion is seen
only in field stars older than 1~Gyr, when we reach stars as cool 
as 5400 K, Be depletion starts at younger ages, between
150 and 600~Myr. 
We have shown that this cannot be explained
by the effect of increasing depth of the convection zone 
(that instead explains
Li depletion), and that an additional source of extra-mixing
is needed also for cool stars. No models including extra-mixing processes
are so far available for stars significantly cooler than the Sun;
our results clearly point toward the necessity of developing such models.

Finally, our Be determination for the two M~67 blue stragglers
does not allow us to draw any definitive conclusion on
their formation scenario; nevertheless, the significant
difference in their Be content suggests that they have
formed by two different processes.
\begin{acknowledgements}
This research has made use of the SIMBAD data base, operated at CDS,
Strasbourg, France. We thank the anonymous referee for the very useful
comments and suggestions.
\end{acknowledgements}
{}
\begin{figure*}
\psfig{figure=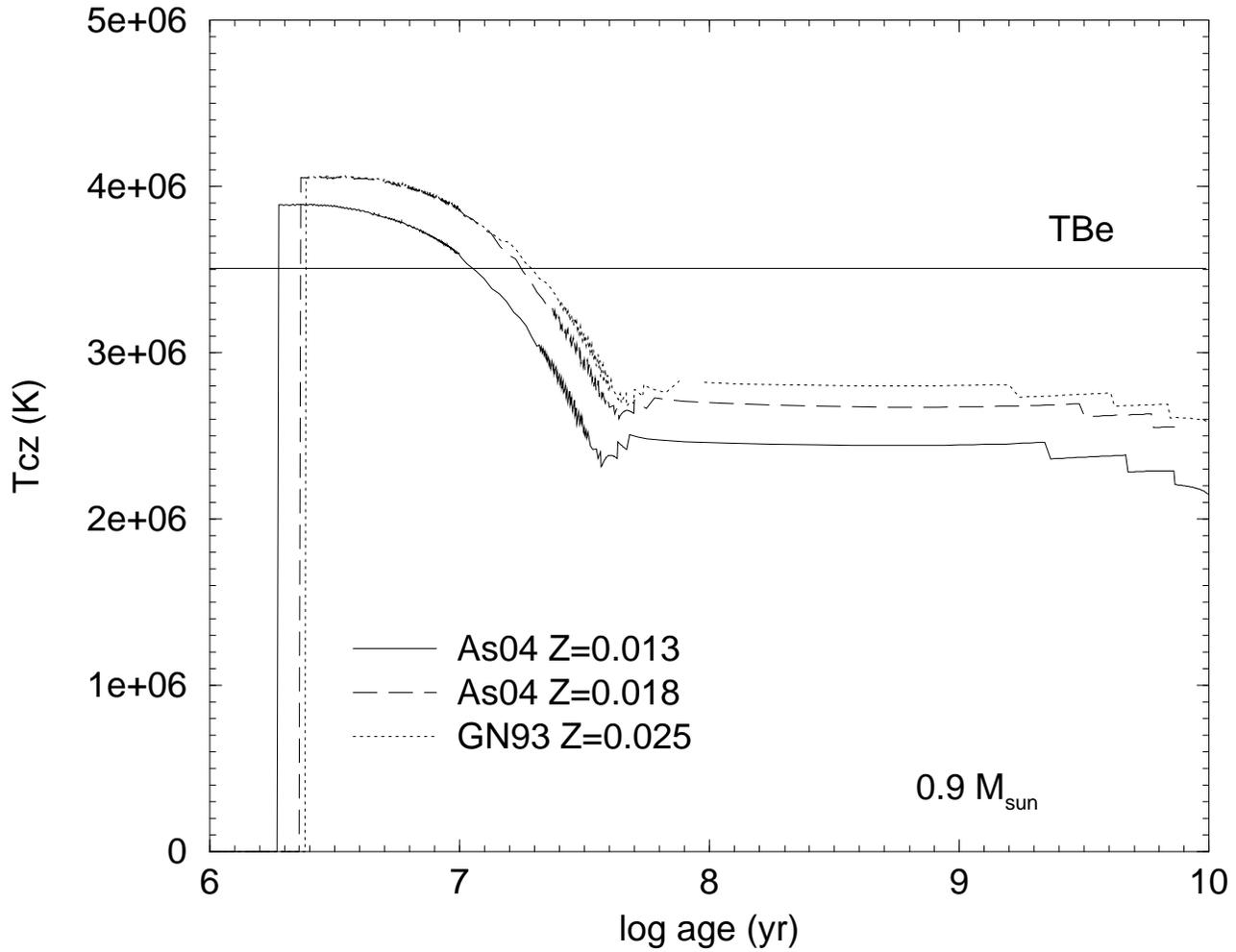, width=17cm}
\caption{Model predictions for the
temperature at the base of the convective zone as a function
of age. The solid horizontal line indicates the temperature necessary
for Be to burn; the solid, dashed, and dotted 
curves show the predictions of the
three different models described in the text.}\label{tcz}
\end{figure*}
\begin{figure*}
\psfig{figure=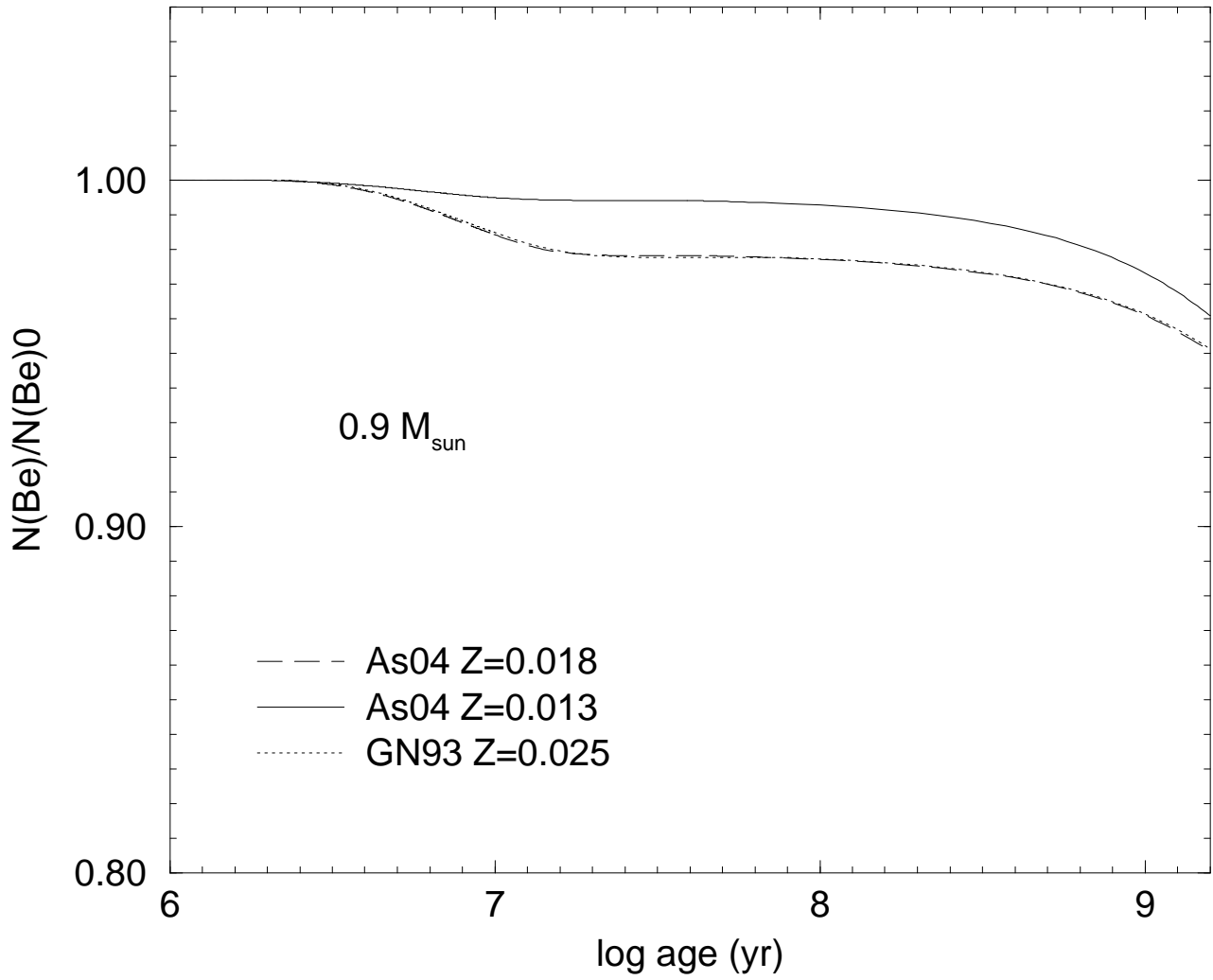, width=17cm}
\caption{Model predictions for the evolution of N(Be)/N(Be)$_0$ as a function
of age for the three different models. 
}\label{nbe_age}
\end{figure*}
\begin{figure*}
\vspace{-3cm}
\psfig{figure=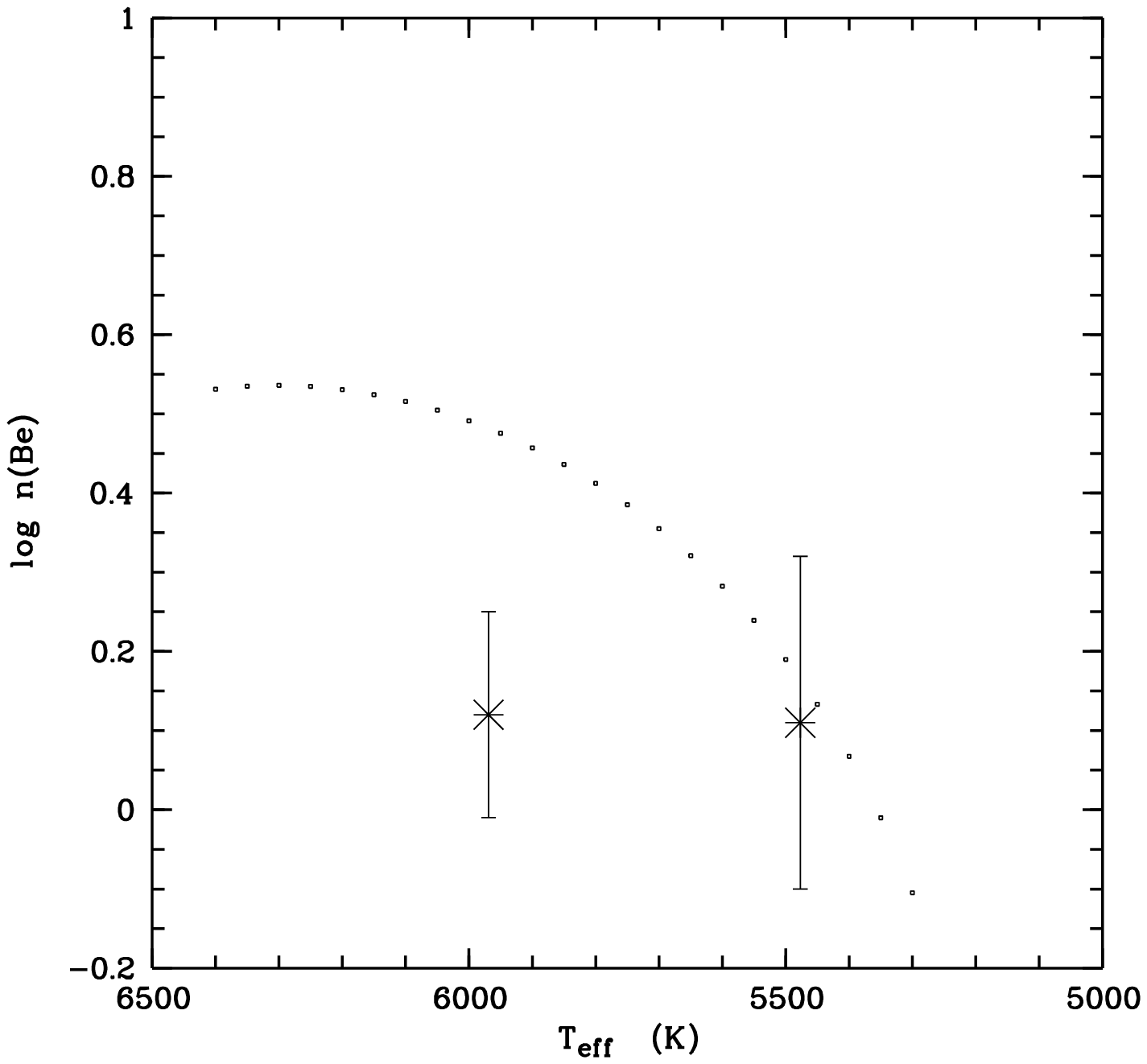, width=17cm, angle=-90}
\caption{Comparison of model predictions of \nbe~for subgiant stars
in M~67 with our measurements for S1034 and S1239. 
The figure is based on Figure~6 in Sills \& Deliyannis (\cite{sd00}),
but the initial Be abundance has been scaled to our value.
With respect to that figure we show only the model including rotationally
driven mixing.}.
 \label{sills}
\end{figure*}

\begin{thebibliography}{}
\bibitem[1989]{ag89}Anders, R., and Grevesse, N. 1989, Geochim. Cosmochim. 
Acta, 53, 197
\bibitem[2004]{asplund04}Asplund, M., Grevesse, N.,
Sauval, A.J., Allende Prieto, C., and Kiselman, D. 2004, A\&A, 417, 751
\bibitem[2005]{asplund05}Asplund, M., Grevesse, N., and Sauval, A.J. 2005,
in Cosmic Abundances as Records of Stellar Evolution and Nucleosynthesis,
eds. T.G. Barnes III and F.N. Bash, ASP Conf. Ser. 336, p.~25
\bibitem[1995]{bala95}Balachandran, S.C. 1995, ApJ, 446, 203
\bibitem[1998]{bb98}Balachandran, S.C., and Bell R.A. 1998,
Nature, 392, 791
\bibitem[1966]{bod66}Bodenheimer, P. 1966, ApJ, 144, 103
\bibitem[1986]{bt86}Boesgaard, A.M., and Tripicco, M.J. 1986, ApJ, 302, L49
\bibitem[1988]{boesetal88}Boesgaard, A.M., Budge, K.G., and Ramsay, M.E.
1988, ApJ, 327
\bibitem[2002]{bk02}Boesgaard, A.M., and King., J.R. 2002, ApJ, 565, 587
\bibitem[2003]{bk03}Boesgaard, A.M., Armengaud, E., and King., J.R. 2003, 
ApJ, 582, 410
\bibitem[2004a]{bk04}Boesgaard, A.M., Armengaud, E., and King., J.R. 2004a, 
ApJ, 605, 864
\bibitem[2004b]{boes04}Boesgaard, A.M., Armengaud, E., and King., J.R., 
Deliyannis, C.P., and Stephens, A. 2004b, ApJ, 613, 1202
\bibitem[1999]{brun}Brun, A. S., Turck-Chi\`eze, S., Zahn, J.-P. 1999,
ASP Conf. Series, 173, p.~173
\bibitem[1995]{chab}Chaboyer, B., Demarque, P., Pinsonneault, M.H. 1995,
ApJ, 441, 865
\bibitem[1999]{ct99}Charbonnel, C., and Talon, S. 1999, A\&A, 351, 635
\bibitem[2005]{ct05}Charbonnel, C., and Talon, S. 2005, Science,  309, 2189
\bibitem[2005]{cp05}Charbonnel, C., and Primas, F. 2005, A\&A, 442, 961
\bibitem[1989]{chieffi}Chieffi, A., and Straniero, O. 1989, ApJS, 71, 47
\bibitem[1975]{chm75}Chmielewski, Y., M\"uller, E.A., and Brault, J.W. 1975,
A\&A, 42, 37
\bibitem[1997]{ciacio97}Ciacio, F., Degl'Innocenti, S.,
and Ricci, B. 1997, A\&AS, 123, 449
\bibitem[2000]{dek}Dekker, H., D' Odorico, S., Kaufer, A., Delabre, B.,
Kotzslowski, H. 2000, Proc. SPIE 4008, 534
\bibitem[1997]{dp97}Deliyannis, C.P., and Pinsonneault, M. 1997, ApJ, 488, 833
\bibitem[1998]{deli98}Deliyannis, C.P., 
Boesgaard, A.M., Stephens, A., et al. 1998, ApJ, 498, L147
\bibitem[1988]{gar88}Garc\'\i a L\'opez, R.J., Rebolo, R., \& Beckmann, J.E.
1988, PASP, 100, 1489
\bibitem[1991]{gs91}Garc\'\i a L\'opez, R.J. and Spruit H.C. 1991, ApJ, 377, 268
\bibitem[1995]{gar95}Garc\'\i a L\'opez, R.J., Rebolo, R., and Perez de Taoro,
M.R. 1995, A\&A, 302, 184
\bibitem[1993]{grevesse93}Grevesse, N., and
Noels, A. 1993 in Origin and Evolution of the Elements, eds.
N. Prantzos, E. Vangioni--Flam, and M. Cass\'e, Cambrdige
University Press, Cambridge, p.~15
\bibitem[1998]{jef98}Jeffries, R.D., James, D.J., and Thurston, M.R. 1998, 
MNRAS, 300, 550
\bibitem[1999]{jon99}Jones, B.F., Fisher, D., and Soderblom, D.R. 1999,
AJ, 117, 330
\bibitem[1993]{kur}Kurucz, R.L. 1993, CD-ROMs \#1, 13, 18
\bibitem[1986]{michaud}Michaud, G. 1986, ApJ, 302, 650
\bibitem[2000]{ms00}Montalb\'an, J., and Schatzmann, E. 2000, A\&A, 354, 943
\bibitem[1981]{nis81}Nissen, P.E. 1981, A\&A, 97, 145
\bibitem[1997]{pas97}Pasquini, L., Randich, S., and Pallavicini, R. 1997,
A\&A, 325, 535
\bibitem[2003]{piau}Piau, L., Randich, S., and Palla, F. 2003, A\&A, 408, 1037
\bibitem[1987]{piletal87}Pilachowski, C.A., Booth, J., Hobbs, L.M.
1987, PASP, 99, 1288
\bibitem[1992]{pins92}Pinsonneault, M.H., Deliyannis, C.P., and Demarque, P.
1992, ApJS, 78, 179
\bibitem[1997]{primas97} Primas, F., Duncan, D.K., Pinsonneault, M.H.,
Deliyannis, C.P., Thorburn, J.A. 1997, ApJ, 480, 78
\bibitem[1991]{pg91}Pritchet, C.J., and Glaspey, J.W. 1991, ApJ, 373, 105
\bibitem[2006]{r_cast} Randich, S. 2006, in Chemical Abundances and Mixing
in Stars in the Milky Way and its Satellites, eds. S. Randich and L. Pasquini,
Springer-Verlag, p.~173
\bibitem[2001]{R01}Randich, S., Pallavicini, R., Meola, G., Stauffer,
J.R., and Balachandran, S. 2001, A\&A, 372, 862
\bibitem[2002]{R02}Randich, S., Primas, F., Pasquini, L., and Pallavicini, 
R. 2002, A\&A, 387, 222 (R02)
\bibitem[2006]{R06}Randich, S., Sestito, P., Primas, F., Pallavicini, R., and 
Pasquini, L. 2006, A\&A, 450, 557
\bibitem[1977]{sand}Sanders, W.L. 1977, A\&AS, 27, 89
\bibitem[2004]{santos}Santos, N.C., Israelian, G.,
Randich, S., Garc\'\i a L\'opez, R.J., Rebolo, R. 2004, A\&A, 425, 1013
\bibitem[2006]{sesti_06}Sestito, P., degl'~Innocenti, S., Prada Moroni, P.,
and Randich, S. 2006, A\&A, 454, 311
\bibitem[1991]{sb91}Schatzman, E., and Baglin, A. 1991, A\&A, 249, 125
\bibitem[2000]{shs00}Shetrone, M.D., and Sandquist, E.L. 2000, AJ, 120, 1913
\bibitem[1997]{sills97}Sills, A., Lombardi, J.C. Jr., Bailyn, C.D., et al.
1997, ApJ, 487, 290
\bibitem[2000]{sd00}Sills, A., and Deliyannis, C.P. 2000, ApJ, 544, 944
\bibitem[1993]{sod93}
Soderblom, D.R., Stauffer, J.R., Hudon, J.D., and Jones, B.F.
1993, ApJS, 85, 313
\bibitem[1987]{spi87}Spite, F., Spite, M., Peterson, R. C.,
Chaffee, F. H., Jr. 1987, A\&A 171, L8
\bibitem[2004]{std04}Steinhauer, A. and Deliyannis, C.P. 2004,
ApJ, 614, L65
\bibitem[1992]{faulk}Swenson, F. J., and Faulkner, J. 1992,
ApJ, 395, 654
\bibitem[1993]{thorb}Thorburn, J.A., Hobbs, L.M., Deliyannis, C.P., and
Pinsonneault, M.H. 1993, ApJ, 415, 150
\bibitem[1994]{thoul}Thoul, A.A., Bahcall, J.N., and
Loeb, A. 1994, ApJ, 421, 828
\bibitem[1952]{vb}van Bueren, H.G. 1952, Bull. astron. Inst. Netherlands, 11, 
385
\end{thebibliography}
\end{document}